# ISSUES AND R&D REQUIRED FOR THE INTENSITY FRONTIER ACCELERATORS*

V.Shiltsev[#], S.Henderson, P.Hurh, I.Kourbanis, V.Lebedev, Fermilab, Batavia, IL 60510, USA

*Abstract*

Operation, upgrade and development of accelerators for Intensity Frontier face formidable challenges in order to satisfy both the near-term and long-term Particle Physics program. Here we discuss key issues and R&D required for the Intensity Frontier accelerators.

## INTENSITY FRONTIER ACCELERATORS

The near-term program of HEP research at the Intensity Frontier continuing throughout this decade includes the long-baseline neutrino experiments and a muon program focused on precision/rare processes. It requires: a) double the beam power capability of the Booster; b) double the beam power capability of the Main Injector; and c) build-out the muon campus infrastructure and capability based on the 8 GeV proton source. The long-term needs of the Intensity Frontier community are expected to be based on the following experiments: a) long-baseline neutrino experiments to unravel neutrino sector, CP-violation, etc.; and b) rare and precision measurements of muons, kaons, neutrons to probe mass-scales beyond LHC.

Both types of experiments will require MW-class beams. The Project-X construction is expected to address these challenges [1]. The Project X represents a modern, flexible, Multi-MW proton accelerator (see Fig.1).

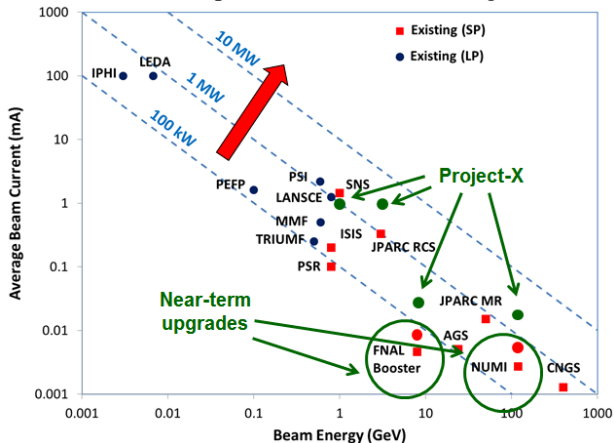

Figure 1: Accelerator beam power landscape.

High power hadron accelerators for the Intensity Frontier have two over-riding design constraints [2]:
*Minimizing beam loss* (typical beam loss requirements for a MW proton beam: < 1 Watt/m, or/and < $1 \times 10^{-4}$ total beam loss), and
*Proper beam structure* (the required formats significantly vary from experiment to experiment – from a quasi-CW for rare particle decays to a single ~2-ns long bunch for MC/NF ). The key challenges to satisfy these constrains include:
• Producing high current, high-quality and high brightness beams with required bunch structure;
• Accelerating high beam currents to high energy with (1) high-duty factors required for high resolution experiments and (2) very low duty factors for neutrino experiments;
• Running multiple experiments in parallel (quasi-simultaneously) required means for beam manipulations on the bunch-by-bunch base;
• Transporting high power beams while maintaining beam loss at a level where routine maintenance is possible;
• Acceleration of beams from keVs to GeVs preserving emittance and minimizing amplitude growth for large-amplitude particles ("beam halo");
• Low-loss extraction of the beams;
• Target systems capable of handling extreme power densities and extreme radiation environments;
• A related challenge is producing intense, high-quality beams of secondary particles (muons, kaons, …).

## ACCELERATOR R&D NEEDED

Below we outline the directions of the accelerator R&D needed to address the above challenges (we omit specific Project X R&D needs which are addressed elsewhere [3], and focus mostly on other important needs, especially, for the Fermilab's future intensity frontier program, which have not been articulated yet):

*High Intensity beam sources:* Radio-frequency quadrupoles (RFQs) are the injector technology of choice at present; the state of the art performance of 100 mA CW demonstrated at LANL/LEDA with 6.7 MeV output energy; is there an alternative technology which could compete with conventional technology?);

*Beam dynamics issues with high intensity beams in existing accelerators (e-cloud, impedances/instabilities):*
<u>RF systems</u> (how to provide RF power for beam acceleration with many-fold higher beam current when the beam induced voltage greatly exceeds RF voltage?);

<u>Space Charge issues</u> (need to understand the Space Charge losses with the higher intensity slipped stacked bunches in MI and Recycler; the need collimators in Recycler (?); realistic Space Charge MI simulations with predictive power that we can compare with data);

SYNERGIA simulations (to include all apertures and magnetic multipoles; compare with beam data);

Electron-Cloud (most effective coating and scrubbing; *in situ* SEY measurements, micro-wave measurements in the Main Injector);

Transition Crossing (evaluation need of a gamma-t jump in MI, transition crossing in Booster);

Injection issues (investigation of painting scenarios and other potential mitigation schemes -  such as use of laser assisted stripping and R&D on rotating injection foil technology - Ultra Nano Crystalline Diamond (UNCD) technology, rotating graphene foil technology);

Loss Control (the need and design of the most efficient collimation schemes)

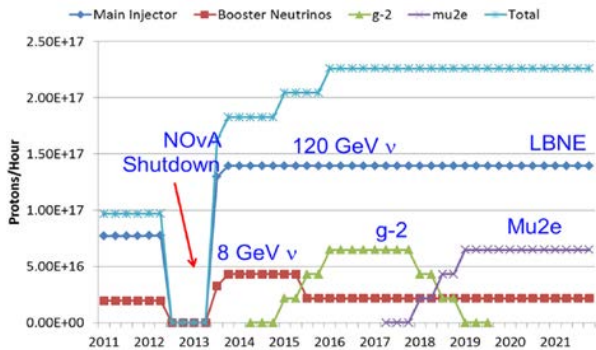

Figure 2: Fermilab's Proton Source proton flux ramp up expectations for the Intensity Frontier experiments.

*Advanced simulation/modeling capabilities for high-intensity beams :* We definitely will need to capitalize on the theoretical developments in beam physics theory and further develop the theory itself. Understanding the coherent instabilities is the corner-stone of the Intensity Frontier developments. Theory of coherent motion of coasting and bunched beams needs to be further studied in various approximations with several complimentary approaches.

Characterization of the loss mechanisms will have significant impact on the entire field of high-intensity accelerators – it is fundamental to loss minimization and control.

Development of a (flexible) beam dynamics framework with fully 3D PIC capabilities (eg on base of SYNERGIA) which will include space-charge & impedance capabilities, both single and multi-bunch effects, single-particle physics with full dynamics and which could run on on desktops, clusters and supercomputers.

Further development of Energy Deposition modeling tools (eg MARS) including: (1) updates related to recent developments in nuclear interaction model;  (2) implementation of polarized particle transport and interaction, (3) recent developments of radiation damage models and transfer matrix algorithms in accelerator material-free regions; and (4) further enhancement of the geometry modules.

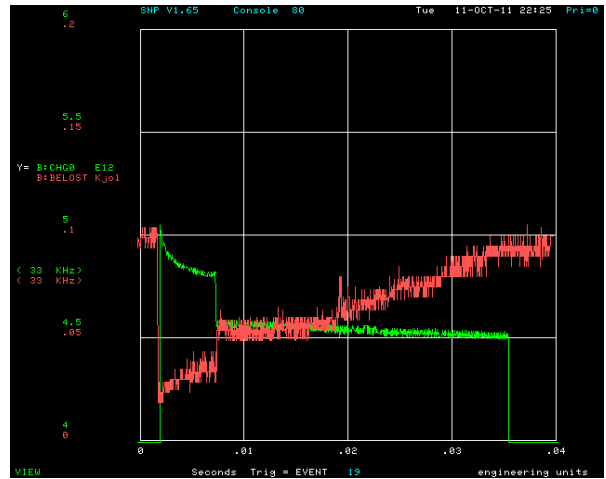

Figure 3: Losses in 8 GeV Fermilab Booster during its 15 Hz cycle occure at injection due to poorly captured beam, at the moment of notch creation (gap for extraction created with a kicker, particles lost in a gradient magnet). slow losses at high-energy (due to optics issues, RF variation), and losses at the transition energy (courtesy of Fermilab's Proton Improvement Plan team).

There is a need in a trustable weak-strong fully symplectic particle tracking codes with a comprehensive set of features for simulation of beam dynamics effects in electron and hadron storage rings, such as dynamics in integrable optics, in presence of space-charge forces and electron compensators, which could also do element-by-element tracking through machine lattice, incorporate FMA methods to assess particle diffusion, etc.

Transfer knowledge from theory and modeling to general R&D (support of the Intensity Frontier-related experiments in existing accelerators and beam test facilities, e.g., ASTA [4]) further into the projects.

*Beam-stripping and beam-chopping with laser-based methods:* The conventional approach to accumulating high proton intensities in a ring is to use multi-turn charge-exchange injection of H- beams, but it is very challenging to apply to >MW class beams, esp. CW beams. Novel laser stripping idea from *Danilov, et. al.* [5] has been demonstrated, but scaling to realistic accelerator parameters is needed. Laser chopping offers very attractive way to form the required beam current structures, and needs to be explored in realistic beam environment.

*Targetry and collimation for Fermilab's future plans, delivery of 2MW beam power on the Long-Baseline Neutrino Experiment (LBNE) target, etc. [6, 7]:*

There is a need in enhanced modeling of beam energy deposition, secondary particle production and collection, radiation damage (DPA), transmutation products (gas production), and residual dose. Also required are advanced simulations of target material (non-linear) response using FEA codes including fracture and/or phase change; exploration/determination radiation damage effects in candidate target and collimator materials at accelerator target facility operating parameters (e.g., via the RaDIATE Collaboration: Radiation Damage In Accelerator Target Environments).Verification and benchmarking of the above mentioned simulation tools and material properties is needed through dedicated testing at beam facilities and autopsy of existing materials.

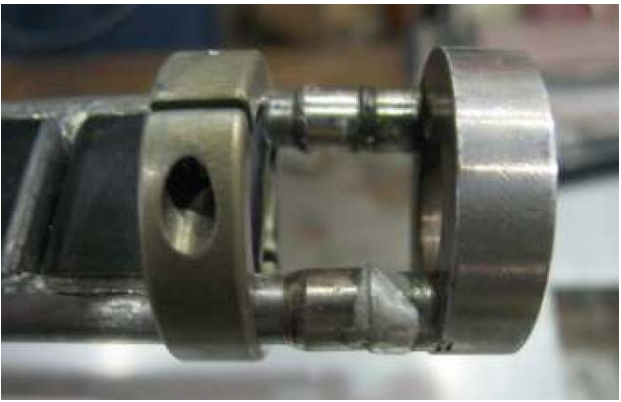

Figure 3: A downstream cooling water turn-around end of NuMI target NT-06 that failed after 4 month of operation at peak beam power of 340kW (toatl of 1.3e20 protons on target, 2011, courtesy J.Hylen, Fermilab)

One also needs to explore and compare high heat flux cooling methods through analysis and testing; to develop novel target and beam window concepts for use in proposed facilities; to develop target facility conceptual designs taking into account full life-cycle and radiation protection issues (remote handling, radioactive component disposal, tritium mitigation, etc.) and diagnostics for use in high radiation environments. We need to continue to engage in the high power targetry community to leverage the collaborative expertise and knowledge base to address the challenges of the High Intensity Frontier.

*New ideas for clean slow extraction:* Slow spills of high intensity beams are needed in particle physics, but beam loss limits intensity. Are there novel ways to cleanly achieve slow spills of high intensity beams? E.g., the resonant extraction - can it be extended to higher orders? Can crystals be used for high-power proton extraction or can electron beams be used to extract protons [8]? Are there other absolutely novel ideas which can employ lasers or high brightness electron beams?

*Revolutionary ideas for Intensity Frontier accelerators:* Integrable Optics [9] promises significant advance for low-loss operation of the IF facilities and must be tested at the IOTA ring at Fermilab's ASTA facility.

Space-charge compensation with either electron columns or electron lenses [10, 11] also has a promise of significant benefit for high intensity beams (due to improved emittance and loss control in high intensity beams) and also needs to be tested at the IOTA in ASTA [12].

*Experimental Characterization and Understanding of High Intensity Beams and Underlying Dynamics (Advanced Instrumentation):* reliable beam instrumentation is needed for precise halo (and emittance) measurements and connection to simulation in a reliable way. Novel diagnostics is required for in-depth understanding of intra-bunch and multi-bunch dynamics (position, modes, tunes, chromaticities, etc.).

One has to note that many of the above listed R&D thrusts are synergetic with the needs of the Accelerator Driven Systems (ADS) [13].

The facilities where the above listed R&D can/will be carried out include existing accelerators (Fermilab's Booster, Recycler and Main Injector) and the ASTA facility. Radiation damage studies for high power targetry may need dedicated facilities for required irradiations and examinations, such as BLIP at BNL for high energy proton irradiations or the proposed UK triple-beam facility, TRITON, for low energy ion irradiations. Verification studies for high power targetry will need dedicated areas at beam facilities capable of delivering intense beam pulses on heavily instrumented test samples, such as HiRadMat at CERN [14-16].